\documentclass{iopjournal}
\usepackage{lmodern}
\usepackage[version=3]{mhchem}
\usepackage{amsmath}
\usepackage{amssymb}
\usepackage{siunitx}
\usepackage{comment}
\usepackage{url}
\usepackage{xcolor}
\usepackage{booktabs}
\usepackage{makecell}

\begin{document}
\articletype{Technical Note} \\
\title{A Practical Guide for Diagnosing Imaginary Phonon Modes in Metal--Organic Frameworks: The Case of MOF-5}
\author{Julia Santana-Andreo$^{1,2}$ and Caterina Cocchi$^{1,2,3}$}\\
\affil{$^1$Institute for Solid-State Theory and Optics, Friedrich-Schiller-Universit\"at Jena, 07743 Jena, Germany}\\
\affil{$^2$Institute of Physics, Carl von Ossietzky Universit\"at Oldenburg, 26129 Oldenburg, Germany}\\
\affil{$^3$Abbe Center of Photonics, Friedrich-Schiller-Universit\"at Jena, 07745, Jena, Germany}\\
\email{julia.santana.andreo@uni-jena.de, caterina.cocchi@uni-jena.de}\\
\begin{abstract}
Assessing the dynamical stability of computationally predicted metal--organic frameworks (MOFs) is essential to distinguish synthetically feasible structures from dynamically unstable ones. However, reliable first-principles phonon calculations on these systems remain challenging: their large, flexible unit cells and soft collective modes make the vibrational spectrum highly sensitive to the numerical settings. Using MOF-5 as a representative case study, we establish a finite-displacement workflow to identify and isolate the origins of imaginary phonon modes. We demonstrate how numerical force convergence thresholds, real-space grid resolutions, symmetry-standardization protocols, and alternative unit-cell representations can qualitatively and spuriously alter the predicted lattice stability. Once numerical noise is confidently excluded, the remaining imaginary modes can be analyzed through mode mapping, with stochastic Monte Carlo symmetry-breaking distortions outlined as a complementary strategy for more complex landscapes. This protocol provides a robust, transferable strategy for the reliable assessment of dynamical stability and lattice vibrations in flexible porous frameworks.
\end{abstract}

\section{Introduction}

Metal--organic frameworks (MOFs) are crystalline porous materials assembled from inorganic nodes and organic molecular linkers that coordinate into extended periodic networks~\cite{yagh+03nature}. This modular architecture underpins the principles of reticular chemistry: by strategically varying the metal coordination chemistry, linker topology, or chemical functionalities, the pore environment and structural response can be precisely tailored~\cite{kalm+18sciadv,chen+22acr}. Consequently, MOFs have emerged as pivotal platforms for compelling applications in carbon capture, gas separation, and energy storage~\cite{chen+22acr,sumi+12cr,kneb+22nnano}.

\begin{figure}
\centering
\includegraphics[width=.9\textwidth]{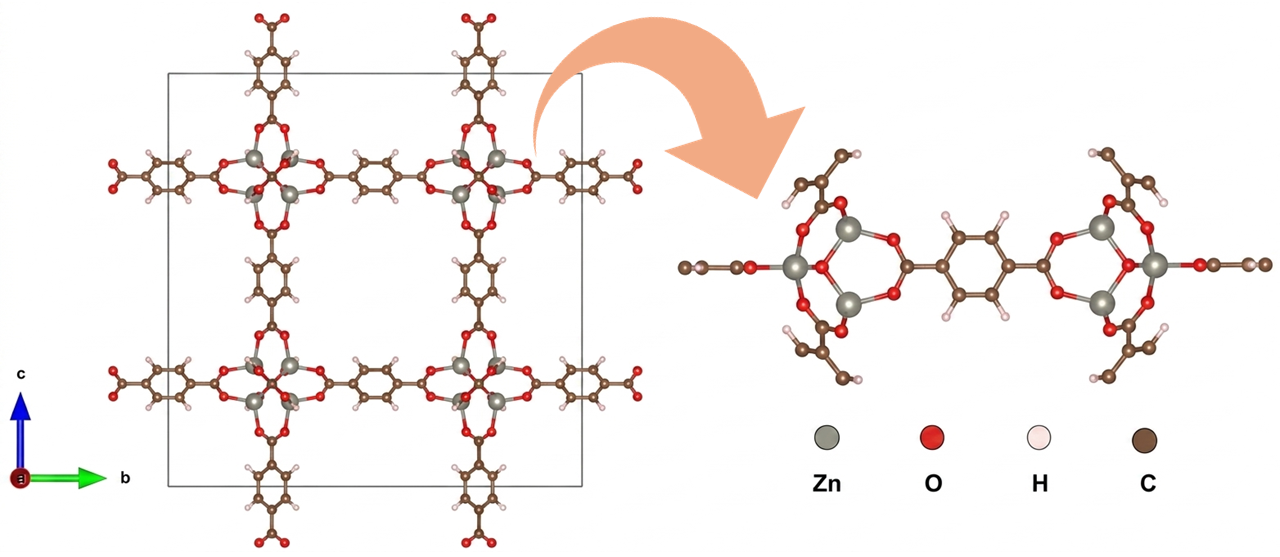}
\caption{Conventional unit cell representation of MOF-5 and zoom-in of the metal--organic linker environment.}
\label{fig:mof_scheme}
\end{figure}

As a prototypical member of this family, MOF-5 serves as a benchmark for both experimental~\cite{li+99nature,edda+00jacs} and computational studies~\cite{lock+13dt,rimm+14pccp,ryde+19adts,edza+25jctc}. Its structure consists of Zn$_4$O nodes connected by 1,4-benzenedicarboxylate (BDC) linkers to form a highly porous cubic framework (Fig.~\ref{fig:mof_scheme})~\cite{li+99nature,edda+00jacs}. Unlike conventional solids, MOF-5 possesses a flexible lattice that hosts soft collective degrees of freedom, such as linker rotations, node distortions, and hinge-like node--linker motions~\cite{lock+13dt,rimm+14pccp,ryde+19adts,andr+20jacs}. These low-energy framework modes govern the response of the materials to temperature and pressure, driving phenomena such as negative thermal expansion and pressure-induced symmetry-lowering distortions~\cite{lock+13dt,rimm+14pccp,ryde+19adts,dubb+07jpcc}. More generally, evaluating accurate phonon dispersion relations is important for assessing the dynamical stability and thermomechanical properties of any porous framework. Since the macroscopic properties of these systems are dictated by low-frequency collective motions rather than localized, high-frequency bond-stretching vibrations, phonons provide the fundamental microscopic link between flexible architecture and engineering performance.

Phonon calculations in MOFs are, however, exceptionally demanding. Their large unit cells, chemically diverse interactions, and shallow potential-energy surfaces (PESs) make their vibrational spectra highly sensitive to the computational setup, often leading to imaginary frequencies. In contrast to dense, covalently bonded semiconductors, where established protocols allow for the direct association of imaginary modes with structural instabilities~\cite{baro+01rmp}, the softer energy landscape of framework materials requires a more nuanced interpretation. Beyond a genuine physical instability, imaginary modes can originate from numerical artifacts: incomplete structural relaxation, under-converged basis sets and $k$-grids, inadequate supercell sizes, or errors introduced during symmetry standardization~\cite{pall+22es,kama+26arxiv,togo+23jpsj}. Identifying and mitigating these numerical issues is essential for assigning physical meaning to a soft mode, particularly in the context of automated high-throughput screening calculations, where system-specific convergence choices are difficult to standardize~\cite{zhu+24npjcmat}.

In this work, we present a finite-displacement workflow for distinguishing spurious numerical artifacts from intrinsic imaginary phonon modes in MOFs, using MOF-5 as a representative case study. We systematically examine how the predicted vibrational spectrum depends on the choice of xc functional, dispersion correction, force-convergence threshold, cutoff parameters, symmetry-standardization tolerance, and unit-cell representation. After isolating these numerical factors, we demonstrate how any persistent soft modes can be analyzed through deterministic mode-mapping calculations, and outline how stochastic Monte Carlo (MC) rattle distortions could provide a complementary symmetry-breaking strategy. The resulting protocol offers practical guidance for achieving reproducible phonon calculations in flexible porous frameworks and clarifies the key technical bottlenecks that remain difficult to transfer directly to automated screening workflows.

\section{Theoretical background}

To establish robust and reproducible guidelines for diagnosing phonon instabilities in MOFs, it is instructive to review the theoretical formalism of the adopted computational workflow. We first summarize the foundations of density-functional theory (DFT), used to evaluate electronic ground states and interatomic forces, and subsequently discuss the harmonic approximation for phonons. By examining the relationship between the dynamical matrix and the curvature of the potential energy surface, we provide the basis for interpreting imaginary frequencies either as physical indicators of structural phase transitions or as artifacts of the numerical setup.

\subsection{Density-functional theory}

The electronic-structure calculations discussed in this work are based on density-functional theory (DFT)~\cite{hohe+64pr}. The key task consists in the solution of the Kohn--Sham (KS) equations~\cite{kohn+65pr}, which in atomic units read:
\begin{equation}
\left[-\frac{1}{2}\nabla^2
+ v_{\mathrm{s}}(\mathbf{r})\right]\psi_i(\mathbf{r})
= \epsilon_i^{\mathrm{KS}}\psi_i(\mathbf{r})
\end{equation}
The effective single-particle potential $v_{\mathrm{s}}(\mathbf{r})$ maps the interacting many-body problem onto a non-interacting reference system and is defined as:
\begin{equation}
v_{\mathrm{s}}(\mathbf{r}) =
v_{\mathrm{ext}}(\mathbf{r}) + v_{\mathrm{H}}(\mathbf{r})
+ v_{\mathrm{xc}}(\mathbf{r})
\end{equation}
where the terms $v_{\mathrm{ext}}$, $v_{\mathrm{H}}$, and $v_{\mathrm{xc}}$ denote the external, Hartree, and xc contributions, respectively. While the external and Hartree potentials can be evaluated exactly from the spatial distribution of the nuclei and the electronic density, the exact form of $v_{\mathrm{xc}}$ is unknown and must be approximated. In this work, we adopt the Perdew--Burke--Ernzerhof (PBE)~\cite{perd+96prl} implementation of the generalized-gradient approximation, and r$^2$SCAN~\cite{furn+20jpcl}, a regularized meta-GGA functional. Both functionals are used together with the Grimme-D3 dispersion correction scheme with Becke--Johnson damping~\cite{grim+10jcp,grim+11jcc} to account for dispersive interactions, with two exceptions. In the force-convergence analysis of Sec.~\ref{sec:sim_params}, r$^2$SCAN is combined with the non-local rVV10 van der Waals functional~\cite{saba+13prb}. In the fragmented-representation test of Sec.~\ref{sec:symmetry}, r$^2$SCAN is used without any dispersion correction.

\subsection{The harmonic approximation}

In the harmonic approximation, the PES is expanded to the second order around an equilibrium configuration, and vibrations are described as normal modes (phonons) with well-defined frequencies and polarization vectors~\cite{born+54dtcl,togo+15scripta}. Let $\mathbf{R}_{l}$ denote the lattice vector of unit cell $l$, and let $\kappa$ label atoms in the primitive cell with masses $M_{\kappa}$. Small displacements from the equilibrium positions are denoted as $u_{l\kappa\alpha}$, where $\alpha \in \{x,y,z\}$. Assuming the structure is relaxed to a stationary point where all the first derivatives vanish, the total energy can be expanded as:
\begin{equation}
E(\{u\}) \approx E_0
+ \frac{1}{2}\sum_{l\kappa\alpha}\sum_{l'\kappa'\beta}
\Phi_{\kappa\alpha,\kappa'\beta}(l,l')\,
u_{l\kappa\alpha}\,u_{l'\kappa'\beta},
\label{eq:harmonic_taylor}
\end{equation}
where $E_0$ is the energy of the equilibrium structure and $\Phi_{\kappa\alpha,\kappa'\beta}(l,l')$ represent the real-space second-order interatomic force constants (IFCs):
\begin{equation}
\Phi_{\kappa\alpha,\kappa'\beta}(l,l')
= \left.\frac{\partial^2 E}
{\partial u_{l\kappa\alpha}\,\partial u_{l'\kappa'\beta}}
\right|_{\{u\}=0}
= -\left.\frac{\partial F_{l\kappa\alpha}}
{\partial u_{l'\kappa'\beta}}\right|_{\{u\}=0}.
\label{eq:force_constants_def}
\end{equation}
Here, $F_{l\kappa\alpha}=-\partial E/\partial u_{l\kappa\alpha}$ is the Hellmann--Feynman force acting on atom $\kappa$ in cell $l$ along the Cartesian direction $\alpha$. Translational invariance dictates the acoustic sum rule,
\begin{equation}
\sum_{l'\kappa'}\Phi_{\kappa\alpha,\kappa'\beta}(l,l')=0,
\label{eq:asr}
\end{equation}
which ensures that the frequencies of the three acoustic phonon branches correctly vanish at the Brillouin-zone center $\Gamma$.

For a periodic solid, the coupled equations of motion decouple into independent blocks for each wavevector $\mathbf{q}$ via the dynamical matrix:
\begin{equation}
D_{\kappa\alpha,\kappa'\beta}(\mathbf{q})
= \frac{1}{\sqrt{M_{\kappa}M_{\kappa'}}}
\sum_{l'}
\Phi_{\kappa\alpha,\kappa'\beta}(0,l')\,
e^{i\mathbf{q}\cdot(\mathbf{R}_{l'}
+\boldsymbol{\tau}_{\kappa'}
-\boldsymbol{\tau}_{\kappa})},
\label{eq:dynamical_matrix}
\end{equation}
where $\boldsymbol{\tau}_{\kappa}$ is the basis vector of atom $\kappa$ in the primitive cell. The phonon frequencies $\omega_s(\mathbf{q})$ and polarization vectors $e^{(s)}_{\kappa\alpha}(\mathbf{q})$ for branch $s$ are obtained by solving the eigenvalue problem:
\begin{equation}
\sum_{\kappa'\beta}
D_{\kappa\alpha,\kappa'\beta}(\mathbf{q})\,
e^{(s)}_{\kappa'\beta}(\mathbf{q})
= \omega_s^2(\mathbf{q})\,
e^{(s)}_{\kappa\alpha}(\mathbf{q}).
\label{eq:eigenproblem}
\end{equation}

A clear physical picture of phonon branches is obtained by introducing the normal-mode coordinate $Q_{s\mathbf{q}}$, allowing the real-space atomic displacements to be expressed as:
\begin{equation}
u_{l\kappa\alpha}(Q_{s\mathbf{q}})
= \frac{1}{\sqrt{M_{\kappa}}}\,
e^{(s)}_{\kappa\alpha}(\mathbf{q})\,
Q_{s\mathbf{q}}\,
e^{i\mathbf{q}\cdot\mathbf{R}_{l}},
\label{eq:mode_coordinate}
\end{equation}
up to an arbitrary phase factor. Within the harmonic approximation, the energy change along $Q_{s\mathbf{q}}$ is purely quadratic:
\begin{equation}
\Delta E(Q_{s\mathbf{q}})
= \frac{1}{2}\,\omega_s^2(\mathbf{q})\,|Q_{s\mathbf{q}}|^2.
\label{eq:mode_energy}
\end{equation}
Therefore, a mode with $\omega_s^2(\mathbf{q}) < 0$ corresponds to negative curvature of the PES along that collective coordinate, causing the calculated harmonic frequency to be formally imaginary. In practice, these imaginary branches indicate either that the reference structure is on a saddle point rather than on a local minimum, or that the numerical setup is insufficiently converged~\cite{pall+22es}. This duality motivates the need for the diagnostic tool proposed in this work.
Throughout this work, we adopt the standard convention of reporting imaginary frequencies as negative values, \textit{i.e.}, a mode with $\omega_s^2(\mathbf{q})<0$ is quoted as $\nu_s(\mathbf{q})= -\sqrt{|\omega_s^2(\mathbf{q})|}/(2\pi)<0$. The most negative eigenvalue of the dynamical matrix $\omega^2_{\min}$ and the corresponding minimum frequency $\nu_{\min}$ provide a compact scalar measure of lattice instability. 

To compute harmonic force constants, one can employ either reciprocal-space density-functional perturbation theory~\cite{baro+01rmp} or real-space finite displacements via the direct method~\cite{parl+97prl,togo+15scripta}. In the finite-displacement approach used here, the IFCs are extracted from the linear force response to a symmetry-reduced set of small atomic displacements generated by \texttt{Phonopy}~\cite{togo+15scripta,togo+23jpsj}. For a displaced configuration $n$, the force difference with respect to the fully relaxed equilibrium structure can be written as:
\begin{equation}
\Delta F_{l\kappa\alpha}^{(n)}
=
-\sum_{l'\kappa'\beta}
\Phi_{\kappa\alpha,\kappa'\beta}(l,l')\,
u_{l'\kappa'\beta}^{(n)}
+\mathcal{O}(\delta^2),
\label{eq:finite_difference}
\end{equation}
where $u_{l'\kappa'\beta}^{(n)}$ denotes the finite displacement applied to atom $\kappa'$ in cell $l'$ along the Cartesian direction $\beta$, $\delta$ is the displacement amplitude and $\Phi_{\kappa\alpha,\kappa'\beta}(l,l')$ represents the second-order IFC matrix. In practice, the system of linear equations generated across all symmetry-inequivalent displaced structures is solved to reconstruct the real-space second-order IFCs, which are then used to build the dynamical matrix via Eq.~\eqref{eq:dynamical_matrix}.

In soft porous materials like MOFs, the low-frequency spectrum is dominated by large-amplitude collective motions of the linkers, including rotations, shears, hinging, and breathing~\cite{rimm+14pccp,andr+20jacs,hoff+22jmca,kuch+19zkri}. Since the PES governing these distortions is exceptionally shallow, the restoring forces are inherently weak. In the harmonic approximation, this flat landscape causes the low-frequency phonons to be critically sensitive to numerical noise. As a direct consequence, (i) the predicted stability of these soft branches can be qualitatively altered by routine settings such as force-convergence thresholds, real-space grid resolutions, supercell sizes, or symmetry constraints, and (ii) marginal variations in geometry, strain, or functional treatment can artificially flip a mode between being stable or unstable. When an imaginary mode persists after a careful screening of these numerical settings, it can be safely classified as a genuine structural instability of the lattice, pointing toward strong lattice anharmonicity and a nearby lower-energy symmetry-broken phase.

\section{Computational methods}
\label{sec:comp_methods}
All first-principles calculations presented in this work are performed using \texttt{CP2K}~\cite{kuhn+20jcp}, version 2024.1. Core electrons are described by Goedecker--Teter--Hutter pseudopotentials~\cite{goed+96prb}, and the KS orbitals are expanded in \texttt{MOLOPT} Gaussian basis sets~\cite{vand-hutt07jcp} with triple-$\zeta$ quality and two polarization functions. Calculations employed the Gaussian and plane-wave (GPW) formalism~\cite{lipp+97mp} implemented in \texttt{QUICKSTEP}, where the KS orbitals are represented by localized Gaussian basis functions, while the electronic density is mapped onto an auxiliary plane-wave grid for the solution of the Poisson equation~\cite{vand+05cpc}. In this framework, the accuracy of the real-space density representation is controlled primarily by the \texttt{CUTOFF} and \texttt{REL\_CUTOFF} parameters: \texttt{CUTOFF} determines the resolution of the finest real-space grid, whereas \texttt{REL\_CUTOFF} controls the assignment of Gaussian products to the multigrid hierarchy. If these parameters are insufficiently converged, the charge density and the resulting atomic forces are affected by grid-integration errors.

Given the large unit-cell size of MOF-5 and its correspondingly small Brillouin zone, all calculations are performed using $\Gamma$-point-only $k$-space sampling. The real-space grid parameters and geometry-optimization thresholds are chosen according to the specific phonon workflow. In the \texttt{CUTOFF}-convergence analysis (Sec.~\ref{sec:sim_params}), the conventional-cell PBE+D3 calculations used \texttt{CUTOFF} values of 2200, 2700, and 3200~Ry at a fixed \texttt{REL\_CUTOFF} of 600~Ry and maximum-force (MF) thresholds of $1.9\times10^{-5}$~Ha/bohr. Conversely, the force-convergence analysis presented in the same section used $2\times2\times2$ primitive-supercell r$^2$SCAN+rVV10 calculations with MF thresholds of $1.9\times10^{-5}$, $1.9\times10^{-6}$, and $1.9\times10^{-7}$~Ha/bohr at fixed \texttt{CUTOFF} and \texttt{REL\_CUTOFF} values of 3200 and 600~Ry, respectively. These stringent numerical settings are necessary because finite-displacement phonon calculations derive IFCs from small force differences, making them exceptionally sensitive to numerical noise.

Since the D3 correction is an analytical, pairwise term depending only on interatomic distances, it does not contribute to the grid-integrated forces. In contrast, the non-local rVV10 kernel is evaluated directly on the electronic density grid alongside the r$^2$SCAN functional, thereby amplifying residual grid-integration noise. We employ r$^2$SCAN+rVV10 for the force-convergence analysis to intentionally expose the worst-case sensitivity of the softest phonon branches to the MF threshold, whereas the grid-independent D3 correction isolates the examined computational parameter in all remaining benchmarks. For the symmetry and cell-representation analysis in Sec.~\ref{sec:symmetry}, we examine three primitive-cell structures obtained  with different setups, all relaxed with a force threshold of $1.9\times10^{-5}$~Ha/bohr: the reference structure (r$^2$SCAN+D3 at 3200/600~Ry \texttt{CUTOFF} and \texttt{REL\_CUTOFF} thresholds), the linker-centered representation (PBE+D3 at 2200/500~Ry), and the fragmented representation (r$^2$SCAN without dispersion correction at 2200/500~Ry). Space-group standardization is performed using \texttt{spglib} as interfaced through \texttt{Phonopy}, with the tolerance explicitly varied within the diagnostic workflow. The expected $Fm\bar{3}m$ symmetry is recovered using \texttt{symprec} values of $5\times10^{-3}$ and $1\times10^{-5}$ for the reference and linker-centered structures, respectively, whereas the fragmented structure requires a looser value of $7\times10^{-3}$.

For the lattice-representation benchmarks in Sec.~\ref{sec:cell_choice}, PBE+D3 calculations are performed using optimized \texttt{CUTOFF}/\texttt{REL\_CUTOFF} pairs of 2200/500~Ry and 3200/600~Ry for the primitive- and conventional-cell representation, respectively. Finally, as detailed in Sec.~\ref{sec:modemapping}, the mode-mapping calculations are performed with the \texttt{ModeMap} code~\cite{skel+17git} using r$^2$SCAN+D3 in the conventional cell. The energy landscape is systematically scanned along the imaginary $A_{2\mathrm{g}}$ mode eigenvector by generating modulated configurations across a normal-mode coordinate range of $-1.5 \leq Q \leq 1.5~\mathrm{amu}^{1/2}\mathrm{\AA}$ with an increment of $0.25~\mathrm{amu}^{1/2}\mathrm{\AA}$. To resolve the resulting double-well potentials without introducing artifacts, the total energy of each unrelaxed, distorted configuration is evaluated via single-point calculations using a highly converged \texttt{CUTOFF}/\texttt{REL\_CUTOFF} pair of 4200/700~Ry and force threshold of $1.9\times10^{-6}$~Ha/bohr.

We emphasize that these tight numerical tolerances are required by the subtle physics of lattice dynamics in soft frameworks. For routine electronic-structure calculations or standard ground-state geometry optimizations, the coarser grid resolutions and looser force thresholds evaluated here are already more than sufficient to achieve excellent convergence in MOF-5, as demonstrated in our previous benchmark study~\cite{edza+25jctc}. The exceptional sensitivity documented in the following sections is a distinct hallmark of finite-displacement workflows, where even minor grid-integration errors or incomplete force relaxations can qualitatively corrupt the delicate curvature of an already shallow potential energy surface.

For the lattice dynamics calculations, the harmonic IFCs are computed with \texttt{Phonopy} using the finite-displacement approach~\cite{togo+15scripta,togo+23jpsj}. Following geometry optimization under tight convergence thresholds to minimize residual stress, the MOF-5 framework is standardized according to its space-group symmetry. Atomic displacements with an amplitude of 0.01~\AA{} are systematically applied to generate the symmetry-inequivalent configurations; for the ideal cubic phase of MOF-5, this mapping requires 19 independent single-point force calculations. The resulting forces are subsequently back-projected to reconstruct the real-space second-order IFCs and obtain the dynamical matrix. Depending on the specific benchmark target, force constants were extracted using either the primitive cell, a $2\times2\times2$ primitive supercell, or the conventional cubic cell. The final phonon dispersion curves are sampled along the standard high-symmetry path for a face-centered cubic Brillouin zone~\cite{sety+10cms}. All structural models and phonon spectra are processed and visualized using \texttt{Python}-based workflows included in the \texttt{aim$^2$dat} package~\cite{sass+24aim2dat}.

For every calculation we extract three scalar diagnostics, collected in Table~\ref{tab:calculation_settings}: the residual acoustic-sum-rule (ASR) violation from Eq.~\eqref{eq:asr}, the smallest eigenvalue $\omega^2_{\min}$ of the dynamical matrix over the sampled $\mathbf{q}$-path, and the corresponding minimum frequency $\nu_{\min}$, with imaginary values reported as negative. We adopt $\nu_{\min}$ as the quantitative stability criterion, regarding a calculation as dynamically stable when $\nu_{\min}\gtrsim-0.01$~THz (reported as $0.00$ in Table~\ref{tab:calculation_settings}). An imaginary mode is classified as a numerical artifact when $\nu_{\min}$ is restored to this window by tightening a single numerical parameter, and as an intrinsic instability only when a finite $\nu_{\min}<0$ persists after the full screening.

\begin{table}[h!]
\caption{\centering Computational settings and adopted parameters for quantitative diagnostics of dynamical stability. For each calculation we report the residual acoustic-sum-rule (ASR) breaking ($\Delta_{\mathrm{ASR}}$), the smallest eigenvalue of the dynamical matrix ($\omega^2_{\min}$), and the corresponding minimum frequency ($\nu_{\min}$); imaginary values are negative, while 0.00 indicates dynamical stability within our numerical resolution ($|\nu_{\min}|\lesssim0.01$~THz).}
\label{tab:calculation_settings}
\footnotesize
\setlength{\tabcolsep}{2.5pt}
\renewcommand{\arraystretch}{1.12}
\begin{tabular}{@{} l l l c c c c c S[table-format=-2.2] S[table-format=-1.2] @{}}
\toprule
Label & $v_{\mathrm{xc}}$ & vdW & \makecell{Cutoff\\(Ry)} & \makecell{MF\\($10^{-5}$~Ha/bohr)} & Cell & Size & \makecell{$\Delta_{\mathrm{ASR}}$\\($10^{-6}$~THz)} & {$\omega^2_{\min}$ (THz$^2$)} & {$\nu_{\min}$ (THz)} \\
\midrule
\multicolumn{10}{@{}l}{\textbf{Section 4.2: Computational Settings \& Convergence}} \\
$\text{CUTOFF}$ & PBE & D3 & 2200/600 & 1.9 & conv & $1\times1\times1$ & 2.37 & -0.59 & -0.77 \\
 & PBE & D3 & 2700/600 & 1.9 & conv & $1\times1\times1$ & 2.35 & 0.00 & 0.00 \\
 & PBE & D3 & 3200/600 & 1.9 & conv & $1\times1\times1$ & 2.37 & 0.00 & 0.00 \\
$\text{Force conv.}$ & r$^2$SCAN & rVV10 & 3200/600 & 1.9 & prim & $2\times2\times2$ & 1.76 & -0.07 & -0.26 \\
 & r$^2$SCAN & rVV10 & 3200/600 & 0.19 & prim & $2\times2\times2$ & 3.66 & -0.10 & -0.31 \\
 & r$^2$SCAN & rVV10 & 3200/600 & 0.019 & prim & $2\times2\times2$ & 3.05 & 0.00 & 0.00 \\
\midrule
\multicolumn{10}{@{}l}{\textbf{Section 4.3: Symmetry Standardization}} \\
Prim.\ reference & r$^2$SCAN & D3 & 3200/600 & 1.9 & prim & $2\times2\times2$ & 2.98 & 0.00 & 0.00 \\
Prim.\ linker & PBE & D3 & 2200/500 & 1.9 & prim & $2\times2\times2$ & 4.59 & 0.00 & 0.00 \\
Prim.\ fragm. & r$^2$SCAN & none & 2200/500 & 1.9 & prim & $2\times2\times2$ & 3.41 & -22.42 & -4.74 \\
\midrule
\multicolumn{10}{@{}l}{\textbf{Section 4.4: Cell Choice \& Supercell Scaling}} \\
Conv.\ cell & PBE & D3 & 3200/600 & 1.9 & conv & $1\times1\times1$ & 2.28 & 0.00 & 0.00 \\
Prim.\ cell & PBE & D3 & 2200/500 & 1.9 & prim & $1\times1\times1$ & 1.42 & -0.07 & -0.27 \\
Prim.\ supercell & PBE & D3 & 2200/500 & 1.9 & prim & $2\times2\times2$ & 1.91 & 0.00 & 0.00 \\
\midrule
\multicolumn{10}{@{}l}{\textbf{Section 4.5: Mode Mapping}} \\
Modemap HSP & r$^2$SCAN & D3 & 4200/700 & 0.19 & conv & $1\times1\times1$ & 2.56 & -0.19 & -0.44 \\
Modemap LSP & r$^2$SCAN & D3 & 4200/700 & 0.19 & conv & $1\times1\times1$ & 2.14 & 0.00 & 0.00 \\
\bottomrule
\end{tabular}
\end{table}

\section{Results and discussion}

\begin{figure}
\centering
\includegraphics[width=0.9\textwidth]{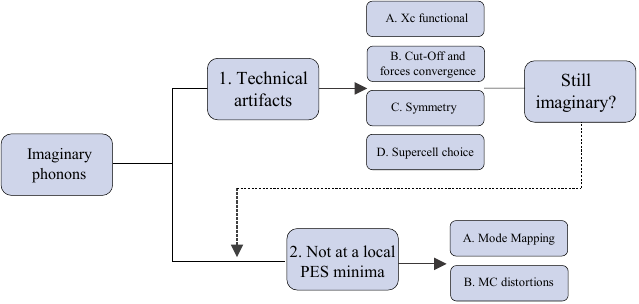}
\caption{Workflow for diagnosing imaginary phonon modes in MOFs.}
\label{fig:imaginary-workflow}
\end{figure}

To distinguish numerical artifacts from intrinsic instabilities in MOFs, we propose a diagnostic workflow using MOF-5 as a case study. The protocol summarized in Fig.~\ref{fig:imaginary-workflow} outlines the recommended validation steps required to obtain reliable phonon dispersions. The process begins by eliminating numerical sources of spurious soft modes, screening parameters such as the xc functional, the grid density, the force convergence thresholds, and the lattice symmetry or supercell settings. Imaginary modes passing these rigorous tests are classified as intrinsic structural instabilities, confirming that the reference configuration is a saddle point rather than a local minimum of the PES. In these cases, mode-mapping techniques and stochastic distortions can be employed to explore the potential energy landscape and identify a stable, symmetry-lowered phase.

The following subsections detail the initial validation branch of this diagnostic protocol. We examine how standard numerical settings can artificially induce or suppress low-frequency imaginary modes in MOF-5, establishing a robust baseline before assigning a physical interpretation to the calculated spectrum. Our analysis begins by assessing the choice of the xc functional and dispersion correction scheme, followed by a detailed investigation into real-space grid resolutions, force tolerances, and spatial symmetry constraints, concluding with the structural impacts of supercell representation. 

\subsection{Exchange-correlation functional and van der Waals forces}
\label{sec:xc}
The sensitivity to the xc functional and the treatment of van der Waals (vdW) interactions stems directly from the softness of the framework lattice. Since vibrational frequencies are governed by the local curvature of the PES, even subtle functional-dependent variations in the equilibrium bond lengths or linker orientations can shift collective branches, such as inter-linker rotations or framework hinging, across the stability threshold. Dispersion forces are particularly critical in this context, as they provide the primary restoring forces for the weak inter-linker couplings that stabilize the lattice against shearing and collapse. Omitting these corrections in DFT calculations typically causes an artificial softening of the interatomic force-constant matrix, leading to spurious imaginary modes.

These physical considerations are supported by recent benchmark results for MOF-5 and its derivatives~\cite{edza+25jctc}. While the qualitative features of phonon dispersion remain robust across different rungs of approximations, ranging from PBE~\cite{perd+96prl}, r$^2$SCAN~\cite{furn+20jpcl}, and even the global hybrid functional PBE0~\cite{adam+99jcp}, the inclusion of vdW corrections is essential for capturing structural nuances and reproducing reliable restoring forces. Interestingly, the numerical settings required to eliminate artificial instabilities vary significantly with the chosen $v_{xc}$. While a maximum-force threshold of $1.9\times10^{-5}$~Ha/bohr is sufficient when using PBE and PBE0, r$^2$SCAN requires substantially stricter settings to stabilize the phonon spectrum~\cite{edza+25jctc}, consistent with the kinetic-energy density dependence of meta-GGA functionals~\cite{sunj+15prl,furn+20jpcl}. Although r$^2$SCAN~\cite{furn+20jpcl} offers improved stability compared to its predecessor SCAN~\cite{sunj+15prl}, it still demands denser real-space grids and tighter force convergence thresholds to minimize numerical noise~\cite{furn+20jpcl,ning+22prb}. In the context of lattice dynamics, this extra sensitivity is magnified: computational settings that lead to fully converged total energies or band structures can still lead to enough force noise to qualitatively distort the softest IFCs, spuriously signaling a structural instability via imaginary modes. This functional dependence motivates the convergence analysis presented in Sec.~\ref{sec:sim_params}.

\subsection{Cutoff and force convergence}
\label{sec:sim_params}
The reliability of finite-displacement phonon calculations critically depends on the numerical quality of the underlying DFT forces. In the adopted GPW formalism (Sec.~\ref{sec:comp_methods}), two primary sources of error can degrade force accuracy: residual forces persisting after geometry optimization and grid-integration errors arising from the real-space representation of the electronic charge density. Because the harmonic IFCs are evaluated as finite differences of these forces [Eq.~\eqref{eq:force_constants_def}], any underlying numerical noise is amplified during the reconstruction of the IFC matrix. This issue is particularly severe for low-frequency modes characteristic of MOFs~\cite{tan+11csrv, redf+19csci,krau+20anie}, where the PES is exceptionally shallow~\cite{parl+97prl, togo+23jpsj}. As detailed in Sec.~\ref{sec:comp_methods}, we adopt for this analysis r$^2$SCAN+rVV10, representing the most numerically demanding setup considered here. Since its non-local kernel shares the real-space grid with the meta-GGA functional, this choice provides a stringent, worst-case probe of how the softest branches respond to the MF threshold.

\begin{figure}
\centering
\includegraphics[width=\textwidth]{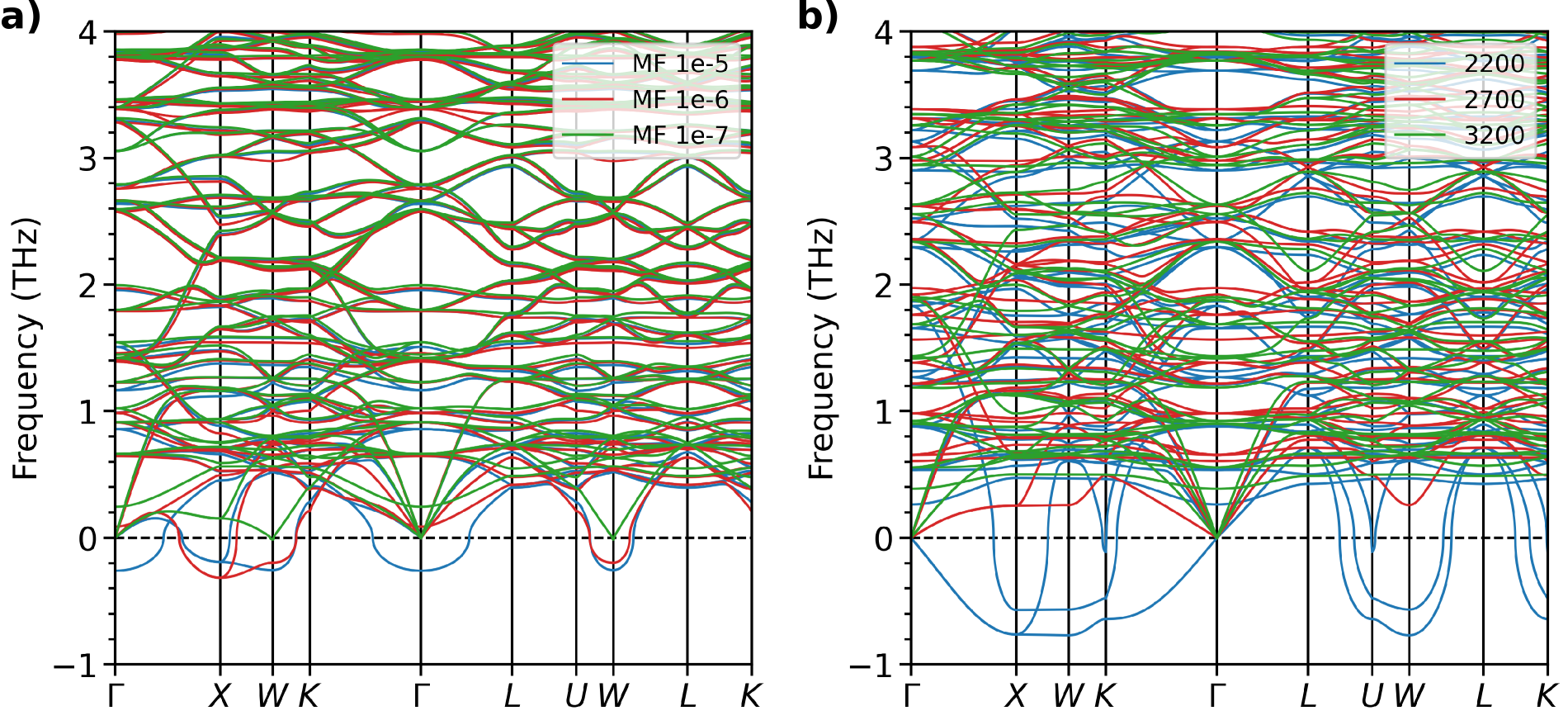}
\caption{Phonon dispersions of MOF-5 calculated (a) using the r$^2$SCAN+rVV10 functional in a primitive $2\times2\times2$ supercell under maximum-force (MF) thresholds of $1.9\times10^{-5}$, $1.9\times10^{-6}$, and $1.9\times10^{-7}$~Ha/bohr, and (b) with PBE+D3 on a conventional unit cell, with \texttt{CUTOFF} values of 2200~Ry, 2700~Ry, and 3200~Ry at fixed \texttt{REL\_CUTOFF} of 600~Ry.}
\label{fig:cutmaxf}
\end{figure}

We evaluated both the maximum-force threshold applied during structural relaxation and the GPW real-space grid parameters. As illustrated in Fig.~\ref{fig:cutmaxf}, variations in either parameter can qualitatively alter the predicted dynamical stability of the MOF-5 lattice. For the r$^2$SCAN+rVV10 calculations performed in the $2\times2\times2$ primitive supercell, tightening the maximum-force threshold from $1.9\times10^{-5}$ to $1.9\times10^{-7}$~Ha/bohr completely eliminates the spurious low-frequency imaginary modes (Fig.~\ref{fig:cutmaxf}a). Quantitatively, the minimum frequency evolves from $\nu_{\min}=-0.26$~THz at the loosest threshold ($1.9\times10^{-5}$~Ha/bohr), through $-0.31$~THz at $1.9\times10^{-6}$~Ha/bohr, to $0.00$~THz at $1.9\times10^{-7}$~Ha/bohr (Table~\ref{tab:calculation_settings}); the non-monotonic intermediate value reflects the residual grid noise that only the tightest threshold fully suppresses. Similarly, isolating the impact of real-space grid density in the conventional-cell PBE+D3 setup (Fig.~\ref{fig:cutmaxf}b) shows that increasing the \texttt{CUTOFF} parameter from 2200~Ry to 3200~Ry (at a fixed \texttt{REL\_CUTOFF} of 600~Ry) is strictly required to suppress unphysical imaginary branches. The coarsest grid (2200~Ry) yields a pronounced imaginary feature with $\nu_{\min}=-0.77$~THz ($\omega^2_{\min}=-0.59$~THz$^2$), spanning acoustic and low-frequency optical regions. This artifact is fully removed ($\nu_{\min}=0.00$~THz) at 2700~Ry and remains stable at 3200~Ry (Table~\ref{tab:calculation_settings}). In both series, the residual ASR violation stays at the $10^{-6}$~THz level irrespective of stability, confirming that ASR compliance alone is not a reliable indicator of numerical convergence for phonon calculations in these materials.

The stringent grid resolution required for this conventional-cell setup does not contradict our previous benchmark study based on PBE~\cite{edza+25jctc}, where a lower \texttt{CUTOFF}/\texttt{REL\_CUTOFF} pair of 2200/500~Ry was found to be sufficient. In that earlier work, performing the calculations within a $2\times2\times2$ primitive supercell fundamentally altered the real-space truncation of longer-range force constants. This variation points to a subtle but tight interplay between real-space grid integration errors and unit-cell representation, which will be discussed in more detail in Sec.~\ref{sec:cell_choice}.

\subsection{Symmetry}
\label{sec:symmetry}
In finite-displacement phonon calculations, crystal symmetry is used to identify equivalent atomic sites and displacement patterns, thereby reducing the number of independent force calculations required to reconstruct the second-order IFCs~\cite{togo+23jpsj}. In the harmonic approximation, the force response on atom $j$ generated by a displacement of atom $i$ is given by:
\begin{equation}
\mathbf{F}_{j} = -\boldsymbol{\Phi}_{ji}\mathbf{u}_{i},
\label{eq:harmonic_force_relation}
\end{equation}
where $\mathbf{F}_{j}$ is the force vector, $\mathbf{u}_{i}$ is the displacement vector, and $\boldsymbol{\Phi}_{ji}$ is the corresponding $3\times3$ Cartesian IFC tensor.
or a space-group operation with Cartesian rotational part $\mathbf{R}$ that maps atoms $i$ and $j$ onto symmetry-equivalent atoms $i'$ and $j'$, respectively, the corresponding IFC block transforms as
\begin{equation}
\boldsymbol{\Phi}_{j'i'} =
\mathbf{R}\boldsymbol{\Phi}_{ji}\mathbf{R}^{\mathrm{T}},
\label{eq:ifc_tensor_transformation}
\end{equation}
where the transpose appears because $\mathbf{R}$ is an orthogonal rotation matrix. This transformation is the Cartesian tensor form of the symmetry mapping used to generate force responses for symmetry-related displacements, avoiding their treatment as independent degrees of freedom~\cite{togo+23jpsj,gan+21cpc}.

In practice, spatial idealization relies on a user-defined numerical tolerance(\texttt{symprec}) to classify equivalent atomic coordinates and lattice vectors~\cite{togo+24stamm}. For flexible MOFs, this standardization step is highly sensitive to tiny residual distortions in the atomic positions. The tolerance factor must therefore be carefully balanced: an excessively loose value may artificially enforce high symmetry on a genuinely broken-symmetry state, while an overly tight threshold fails to identify the actual space group due to numerical noise. Ultimately, the tolerance required to recover the nominal symmetry is itself a diagnostic metric. Since this threshold can slightly shift by small factors between equally valid relaxations, we do not treat the \texttt{symprec} value alone as a conclusive parameter. An under-converged geometry is identified by the combination of a looser symmetry tolerance and a simultaneous degradation of $\nu_{\min}$.

\begin{figure}[h!]
\centering
\includegraphics[width=\textwidth]{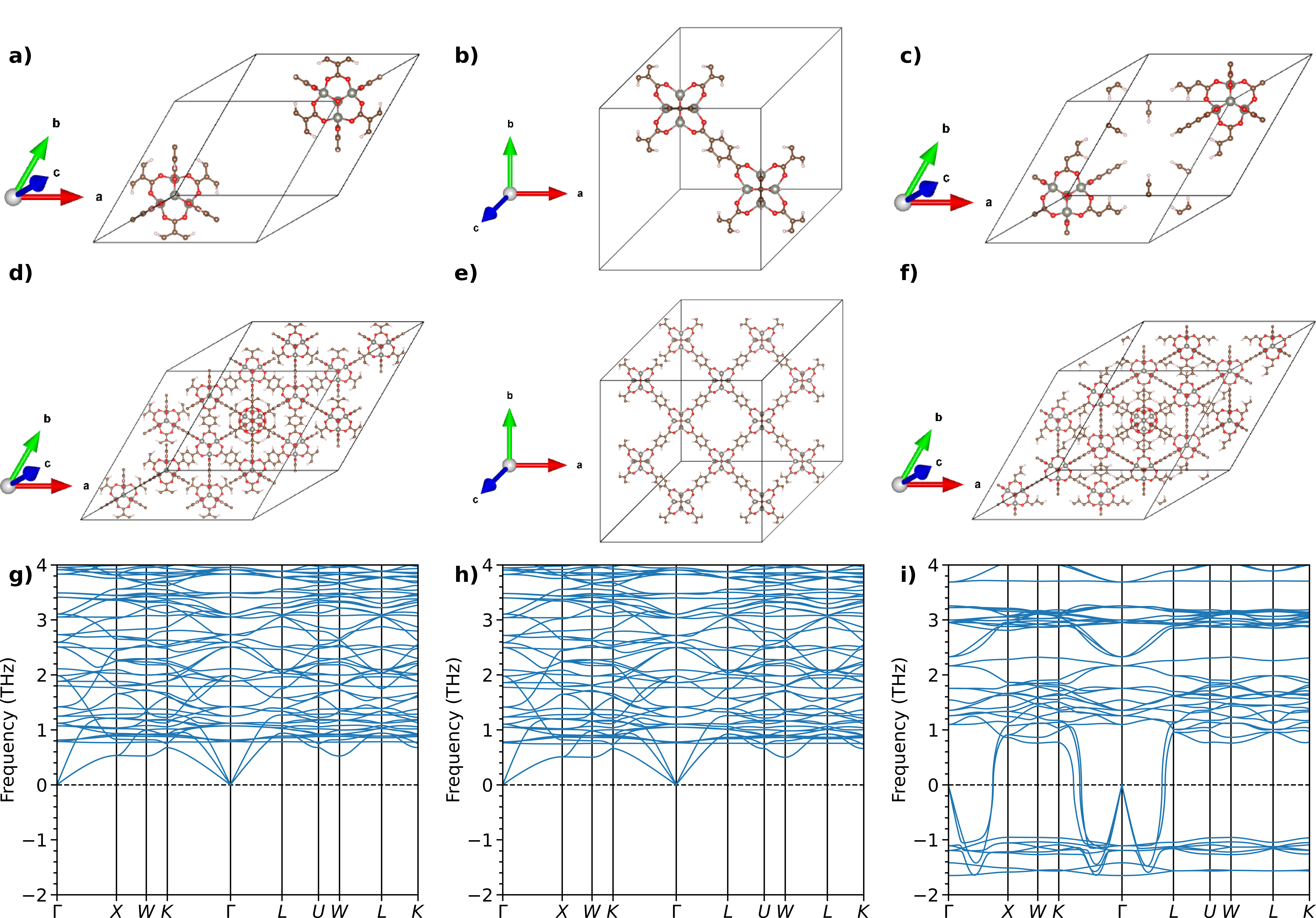}
\caption{Primitive-cell representations of MOF-5 used to assess the symmetry-standardization procedure:(a) reference representation (r$^2$SCAN+D3, 3200/600~Ry), (b) origin-shifted, linker-centered representation (PBE+D3, 2200/500~Ry), and (c) fragmented representation (r$^2$SCAN without dispersion correction, 2200/500~Ry).} Panels (d–f) illustrate the corresponding $2\times2\times2$ supercell expansions, and (g–i) display the resulting harmonic phonon dispersion relations.
\label{fig:sym_prim}
\end{figure}

To assess the impact of this symmetry-standardization step, we examine three primitive-cell representations of MOF-5 (Fig.~\ref{fig:sym_prim}) obtained with the computational settings detailed in Sec.~\ref{sec:comp_methods} and Table~\ref{tab:calculation_settings}. The purpose of this comparison is not to re-assess the individual roles of the exchange-correlation functional or the dispersion scheme, which were systematically benchmarked in Ref.~\cite{edza+25jctc}, but to establish how the quality of the underlying relaxation propagates into the standardization step: specifically, which \texttt{symprec} is required to recover the nominal $Fm\bar{3}m$ symmetry, which coordinate wrapping is selected, and whether the resulting dispersion is dynamically stable. The reference structure shown in Fig.~\ref{fig:sym_prim}a recovers $Fm\bar{3}m$ at \texttt{symprec}~$=5\times10^{-3}$. Expanding this cell into a $2\times2\times2$ supercell (Fig.~\ref{fig:sym_prim}d) preserves the continuous framework connectivity across the periodic boundaries, yielding a dynamically stable phonon dispersion free of imaginary modes ( $\nu_{\min}=0.00$~THz), see Fig.~\ref{fig:sym_prim}g and Table~\ref{tab:calculation_settings}.

A second, independently relaxed structure yields the alternative primitive-cell representation shown in Fig.~\ref{fig:sym_prim}b (PBE+D3, 2200/500~Ry, see Table~\ref{tab:calculation_settings}), which again recovers the $Fm\bar{3}m$ space-group symmetry with a tolerance of $1\times10^{-5}$. Here, the symmetry-standardization algorithm selects a different translation wrapping, centering the organic linkers within the visualization boundaries. Expanding this cell into a $2\times2\times2$ supercell (Fig.~\ref{fig:sym_prim}e) creates a visually different pattern from the reference supercell (Fig.~\ref{fig:sym_prim}d). Remarkably, this periodic wrapping does not affect lattice dynamics: atomic coordinates related by lattice-vector translations describe the identical crystal. Consistently, this linker-centered representation yields a stable dispersion ($\nu_{\min}=0.00$~THz), see Fig.~\ref{fig:sym_prim}h.

A qualitatively different outcome is obtained for the third structure, relaxed with r$^2$SCAN at 2200/500~Ry without dispersion correction (Table~\ref{tab:calculation_settings}). This resolution is adequate for PBE but under-converged for the meta-GGA functional~\cite{edza+25jctc}. Consequently, $Fm\bar{3}m$ symmetry is detected only when \texttt{symprec} is loosened to $7\times10^{-3}$. Symmetrization under these conditions results in a visibly fragmented primitive-cell representation (Fig.~\ref{fig:sym_prim}c). The resulting phonon dispersion is extremely unstable with $\nu_{\min}=-4.74$~THz ($\omega^2_{\min}=-22.42$~THz$^2$): such a value is roughly one order of magnitude larger than the sub-THz spurious features caused by grid or force under-convergence alone (Fig.~\ref{fig:sym_prim}i and Table~\ref{tab:calculation_settings}). 

\begin{figure}[h!]
\centering
\includegraphics[width=0.9\textwidth]{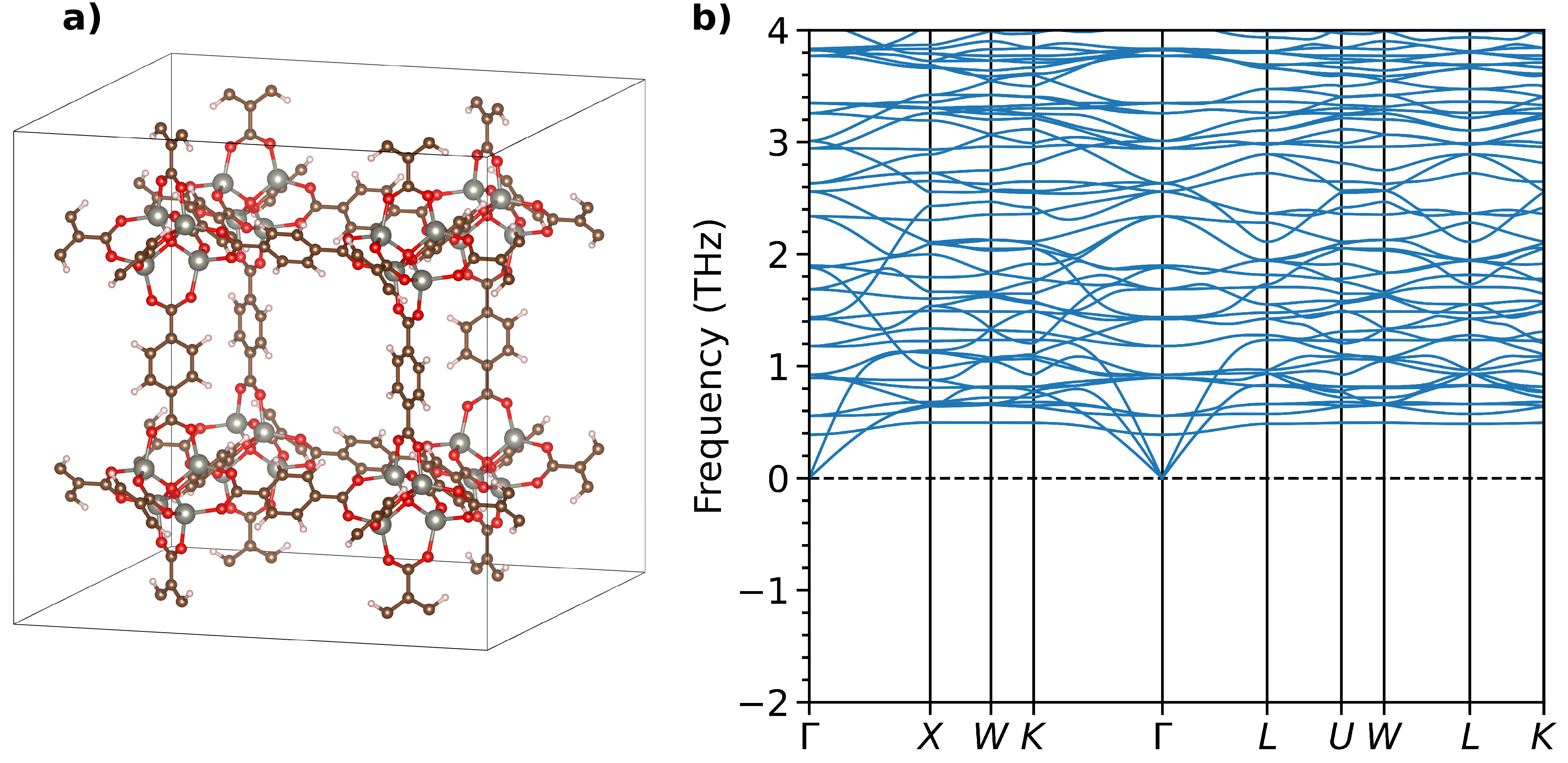}
\caption{Conventional cubic cell reference used for the lattice dynamics of MOF-5. (a) Symmetry-standardized conventional structure and (b) its corresponding harmonic phonon dispersion relation.}
\label{fig:sym_conv}
\end{figure}

It is worth stressing that this fragmentation is an artifact of the representation rather than an actual structural issue. A minimum-image bond-graph analysis confirms that all three primitive cells are single-connected periodic networks with identical coordination numbers and mutually isomorphic bond graphs, with bond lengths in the fragmented cell matching the reference to within $0.005$~\AA{} (further details are reported in Sec.~S1 of the Supporting Information). The residual forces left by the coarse meta-GGA grid require a looser \texttt{symprec} threshold during standardization, causing an unusual coordinate wrapping to be selected. Because this anomalous wrapping is visible immediately after standardization, it acts as an effective primary visual check. While alternate cell wrappings are harmless in converged structures (as proven by the linker-centered cell), a visually fragmented framework signals that the real-space grid resolution is insufficient for lattice dynamics.  To remove this ambiguity, the conventional cubic cell provides a superior alternative for phonon calculations. It consistently recovers a stable reference structure across various tolerances and coordinate representations (Fig.~\ref{fig:sym_conv}), establishing a reproducible foundation for subsequent supercell expansion and yielding a well-behaved harmonic spectrum.

\subsection{Choice of the unit cell and supercell}
\label{sec:cell_choice}
In the finite-displacement formalism implemented in \texttt{Phonopy}, IFCs are evaluated in a periodic supercell, effectively truncating interactions beyond the minimum-image distance. This truncation propagates directly into the Fourier interpolation of the dynamical matrix [Eq.~\eqref{eq:dynamical_matrix}], primarily affecting the lowest-frequency acoustic and optical branches where the restoring forces are weakest~\cite{togo+23jpsj}. The issue can be particularly pronounced in porous frameworks, whose soft collective modes often involve weak but spatially extended couplings~\cite{kuch+19zkri,hoff+22jmca}.

\begin{figure}[h!]
\centering
\includegraphics[width=\textwidth]{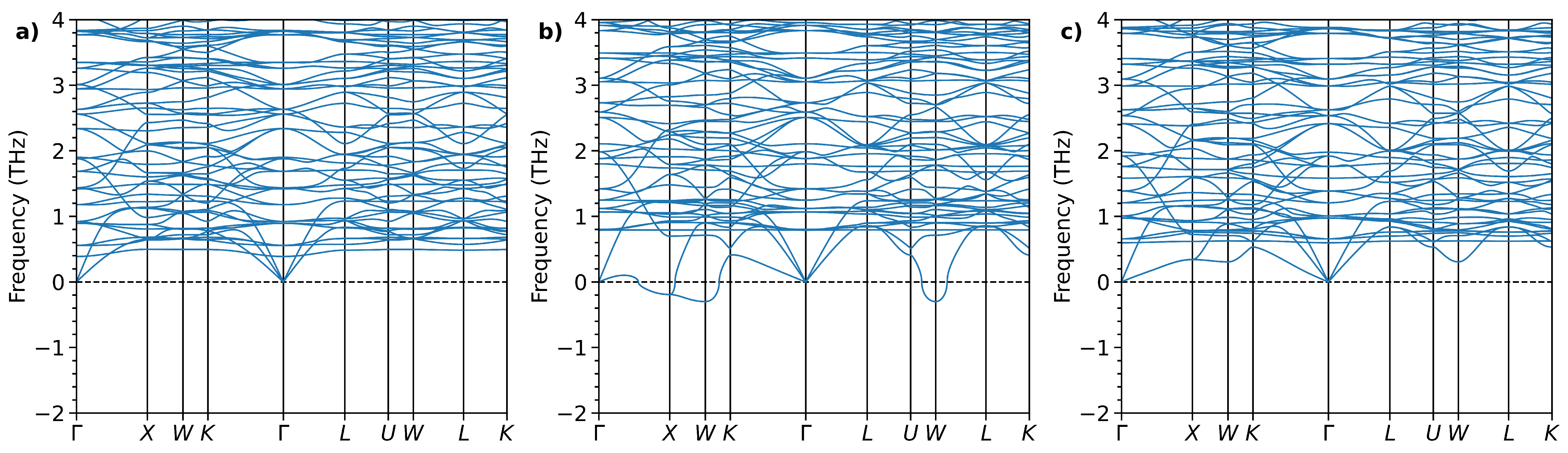}
\caption{Effect of unit-cell and supercell representation on the harmonic phonon dispersion of MOF-5. Results of calculations performed for (a) the conventional unit cell, (b) the primitive unit cell, and (c) the primitive $2\times2\times2$ supercell are shown.}
\label{fig:cell_choice_mof5}
\end{figure}

To mitigate these artifacts, larger supercell expansions are conveniently employed, although demanding a substantially steeper computational cost. For the cubic phase of MOF-5, the primitive rhombohedral cell contains 106 atoms, the conventional cubic cell accommodates 424 atoms (corresponding to a fourfold expansion of the primitive volume), and the $2\times2\times2$ primitive supercell hosts 848 atoms. To isolate the impact of the cell selection from grid-aliasing errors, the phonon dispersion curves displayed in Fig.~\ref{fig:cell_choice_mof5} are obtained using \texttt{CUTOFF}/\texttt{REL\_CUTOFF} pairs specifically converged for each setting: 2200/500~Ry for the primitive-cell calculations and 3200/600~Ry for the conventional-cell calculations. As established in Sec.~\ref{sec:sim_params}, the \texttt{CUTOFF} required to stabilize soft branches is highly sensitive to the spatial representation of the lattice.

As illustrated in Fig.~\ref{fig:cell_choice_mof5}b, the unexpanded primitive-cell calculation exhibits small, spurious imaginary frequencies near the zone center $\Gamma$. This residual instability is quantitatively mild, $\nu_{\min}=-0.27$~THz ($\omega^2_{\min}=-0.07$~THz$^2$). These unstable branches completely disappear when using the conventional cubic cell (Fig.~\ref{fig:cell_choice_mof5}a) or the $2\times2\times2$ primitive supercell representation (Fig.~\ref{fig:cell_choice_mof5}c), both of which reach $\nu_{\min}=0.00$~THz (Table~\ref{tab:calculation_settings}). This trend provides a powerful diagnostic indicator: if imaginary modes vanish as the real-space cutoff radius of the IFCs is extended, they can be confidently classified as numerical artifacts rather than signatures of an intrinsic structural instability~\cite{togo+23jpsj}.

This behavior highlights the distinct real-space ranges offered by different lattice representations. Since the primitive cell is the minimal repeating unit of the lattice, it provides a very short real-space cutoff radius for force-constant evaluation, which is often insufficient to capture the long-range collective motions characteristic of MOFs. In contrast, the conventional cubic cell encompasses four times the volume and spans a significantly larger real-space interaction radius, allowing the IFCs to decay naturally over longer physical distances and thus providing a robust approximation of the vibrational spectrum that matches the accuracy of a full $2\times2\times2$ primitive supercell.

The practical advantages of using the conventional cell over a primitive supercell are substantial. For MOF-5, both the conventional-cell and the $2\times2\times2$ primitive supercell give rise to 19 symmetry-independent displacement configurations. However, because each conventional-cell force calculation scales with half of the atoms, it is approximately 2.5 times faster. As a result, the entire workflow is completed in roughly 40\% of the total wall time required by the supercell approach~\cite{edza+25jctc}. Given this superior numerical efficiency, combined with its inherent robustness against the symmetry-standardization issues discussed in Sec.~\ref{sec:symmetry}, we strongly recommend the conventional unit cell as the standard reference for production-level lattice dynamics calculations in cubic MOFs.

\subsection{Mode mapping of soft phonon instabilities}
\label{sec:modemapping}
Once numerical artifacts have been ruled out, the persistence of an imaginary phonon mode indicates that the reference structure does not correspond to a true local minimum on the PES, but rather a saddle point or a local maximum along that specific collective distortion coordinate~\cite{pall+22es}. Mode-mapping techniques can be used to track the total energy landscape along the unstable normal coordinate, identifying the symmetry-lowered minimum connected to the instability. This procedure involves the generation of a set of distorted structures by displacing atomic coordinates along the eigenvectors of the unstable mode, subsequently evaluating the total energy as a function of the mass-weighted normal coordinate $Q\equiv Q_{s\mathbf{q}}$ introduced in Eq.~\eqref{eq:mode_coordinate}. 

In the harmonic limit, the energy profile follows the quadratic relation outlined in Eq.~\eqref{eq:mode_energy}. For an unstable branch where $\omega^2<0$, the local curvature at $Q=0$ is inherently negative. Stabilization at finite displacements is recovered through higher-order anharmonic terms, which are typically associated with a characteristic double-well potential~\cite{skel+16prl,whal+16prb}. While the upcoming discussion focuses on a one-dimensional scan of the unstable $A_{2g}$ soft mode, the same strategy can be extended to multidimensional configuration spaces, $\Delta E(Q_1,Q_2)$, to resolve complex structural transitions where the global energy minimum is reached via a coupled superposition of multiple unstable branches.

\begin{figure}
\centering
\includegraphics[width=\textwidth]{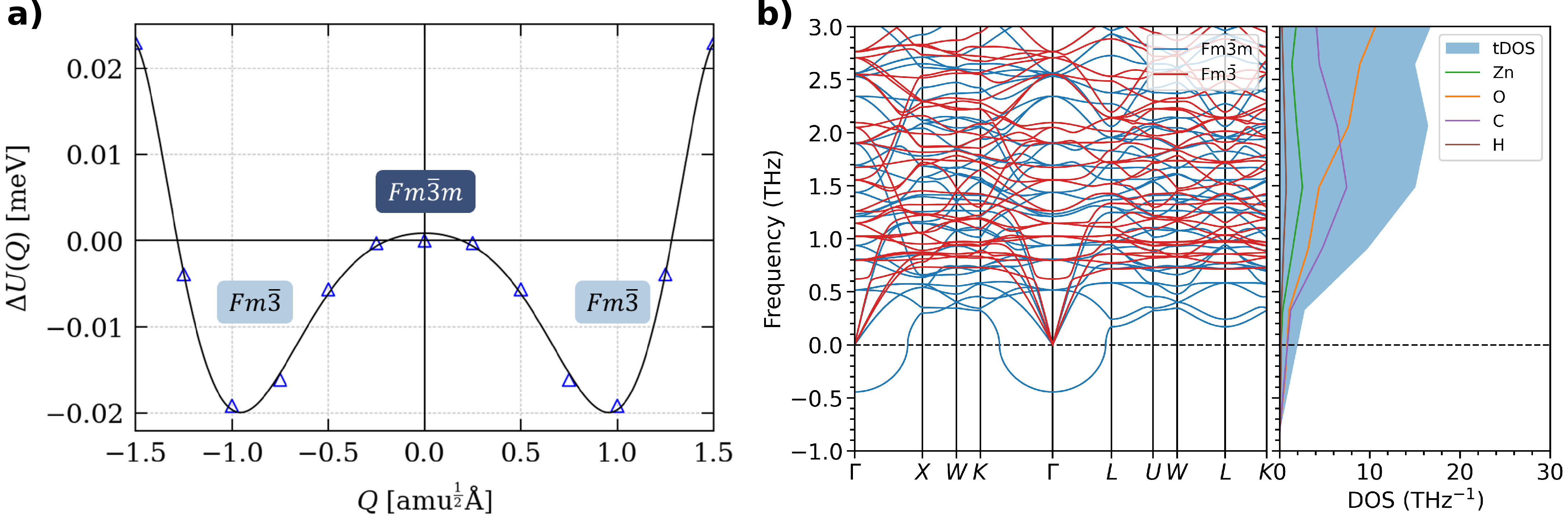}
\caption{Mode mapping of the unstable $A_{2g}$ phonon mode in MOF-5. (a) One-dimensional potential-energy scan along the unstable normal coordinate $Q$. Triangles indicate total energies computed from single-point DFT calculations using the r$^2$SCAN+D3 functional, referenced to the high-symmetry $Fm\bar{3}m$ structure at $Q=0$ and given per conventional unit cell; the solid line is a fit to an even polynomial in $Q$ and the dashed line the harmonic potential implied by the calculated imaginary frequency of $-0.44$~THz. (b) Phonon dispersion and atom-projected phonon density of states for the high-symmetry $Fm\bar{3}m$ reference structure (blue), alongside the phonon dispersion of the symmetry-lowered $Fm\bar{3}$ phase (red).}
\label{fig:modemap_mof5}
\end{figure}

The one-dimensional energy profile $\Delta E(Q)$, computed for the soft mode in MOF-5 using the \texttt{ModeMap} code~\cite{skel+17git} (r$^2$SCAN+D3 calculation in the conventional unit cell with a \texttt{CUTOFF}/\texttt{REL\_CUTOFF} pair of 4200/700~Ry and force threshold of $1.9\times10^{-6}$~Ha/bohr) reveals the physical origin of this structural instability (Fig.~\ref{fig:modemap_mof5}a). At this setup, the high-symmetry reference structure (HSP) retains a genuine soft mode with $\nu_{\min}=-0.44$~THz ($\omega^2_{\min}=-0.19$~THz$^2$), while the mode-mapped, lower-symmetry structure (LSP) is dynamically stable with $\nu_{\min}=0.00$~THz (Table~\ref{tab:calculation_settings}). In contrast to the artifacts of Secs.~\ref{sec:sim_params} and~\ref{sec:cell_choice}, this residual imaginary frequency does not disappear upon further tightening of the numerical settings, and is thus classified as an intrinsic instability. At $Q=0$, corresponding to the undistorted $Fm\bar{3}m$ structure, the system sits at the saddle point of a shallow symmetric double-well potential. Finite distortions along either direction ($\pm Q$) lead the framework into equivalent energy minima associated with a reduced $Fm\bar{3}$ space-group symmetry. This behavior represents the expected fingerprint of a displacive instability.

The potential wells are shallow but well resolved. Fitting the computed profile to an even polynomial in $Q$ (Table~S2) places the minima at $Q=\pm0.96~\mathrm{amu}^{1/2}\mathrm{\AA}$, located $0.54$~meV per conventional unit cell ($1.3~\mu$eV per atom) below the HSP. This energy difference is well above numerical noise: the scatter between symmetry-equivalent $\pm Q$ pairs does not exceed $2.5\times10^{-8}$~meV across the scan, and tightening the SCF threshold from $10^{-7}$ to $10^{-9}$~Ha shifts $\Delta E$ by merely $4\times10^{-6}$~meV (Table~S1). The curvature of the fitted potential at the origin corresponds to a frequency of $-0.46$~THz, within 4.5\% of the $-0.44$~THz obtained by diagonalizing the dynamical matrix (dashed line in Fig.~\ref{fig:modemap_mof5}a). Beyond $|Q|\approx0.3~\mathrm{amu}^{1/2}\mathrm{\AA}$, the profile departs from the parabolic approximation, and the resulting anharmonic stabilization generates the double well. The sub-meV depth of these local wells explains why these low-frequency phonons are so sensitive to numerical details: the framework resides in immediate energetic proximity to competing structural configurations.

The calculated dispersions obtained for both the high-symmetry $Fm\bar{3}m$ reference structure and the symmetry-lowered $Fm\bar{3}$ phase (Fig.~\ref{fig:modemap_mof5}b) confirm this interpretation. The structure at the potential minimum was fully relaxed using the same settings as the HSP and retains $Fm\bar{3}$ symmetry, confirming that the identified minimum is a genuine local minimum of the PES. Structural changes are restricted to the low-frequency domain: the imaginary branches characteristic of the $Fm\bar{3}m$ lattice disappear upon transitioning to the $Fm\bar{3}$ structure. This behavior demonstrates that the instability reflects a highly localized, collective distortion that lowers spatial symmetry to stabilize the framework, rather than a global breakdown of the harmonic approximation. In the low-frequency regime, the atom-projected phonon density of states is dominated by O and C contributions, accompanied by secondary Zn motion; the negligible H contribution confirms that this soft branch corresponds to the collective deformation of the node--linker network rather than local ligand librations.

Given the shallowness of the wells, it is essential to consider whether the static $Fm\bar{3}$ distortion survives nuclear quantum motion. Solving the one-dimensional Schr\"odinger equation within the fitted potential places the vibrational ground state approximately $2$~meV above the barrier at $Q=0$, ensuring $\langle Q\rangle=0$ even at $T=0$. Furthermore, the $0.54~\mathrm{meV}$ barrier height is almost two orders of magnitude smaller than $k_\mathrm{B}T$ at room temperature. Hence, the time- and ensemble-averaged structure retains the full $Fm\bar{3}m$ symmetry, consistent with the experimental cubic phase. The $Fm\bar{3}$ minima are thus best read as a feature of the potential-energy landscape at zero temperature rather than as static structural distortions, quantifying how close MOF-5 resides to the threshold of dynamical stability.

Importantly, the absence of such an instability in the calculations performed on the primitive $2\times2\times2$ supercell does not imply that supercell expansion artificially stabilizes the framework. Instead, it reveals that the soft mode lies in the immediate proximity of the stability boundary, rendering it exceptionally susceptible to tiny variations in reference geometry, symmetry treatment, real-space force containment, and numerical convergence tolerances. This extreme sensitivity is consistent with the quasi-harmonic analysis by Ryder \textit{et al.}~\cite{ryde+19adts}, who demonstrated that isotropic lattice compression softens the low-energy $A_{2g}$ mode in MOF-5, inducing a spontaneous $Fm\bar{3}m \to Fm\bar{3}$ transition. A related pressure-dependent mechanism was identified by Bhogra and Waghmare~\cite{bhog-wagh21jpcc}, who linked hydrostatic compression to nearly dispersionless mid-frequency unstable modes that destabilize the Zn$_4$O secondary building units, providing a microscopic route for amorphization at ultra-low pressures. In the study by Ryder \textit{et al.}~\cite{ryde+19adts}, this instability appeared only under compression because their reference structure, obtained with a hybrid functional, was slightly more expanded than ours computed with r$^2$SCAN, highlighting the delicate interplay between functional-dependent lattice constants and intrinsic mechanical soft modes in flexible frameworks.

\subsection{Stochastic symmetry breaking via Monte Carlo rattle distortions}
\label{sec:mc_rattle}
Stochastic symmetry-breaking protocols provide a robust alternative for mapping complex energy landscapes where multiple imaginary modes coexist and interact. This approach is implemented via the MC-rattle procedure in \texttt{hiphive}~\cite{erik+19adts}, where atoms are randomly displaced from their ideal lattice positions subject to a hard-sphere distance filter that eliminates unphysical overlap. A physically motivated related alternative is the phonon-rattle scheme~\cite{erik+19adts,west+06prl}, which constructs initial perturbations by superposing harmonic normal modes with randomized amplitudes and phases scaled to a target temperature. By biasing displacements along thermally accessible collective coordinates while respecting atomic mass and local bonding constraints, phonon rattling reduces reliance on empirical displacement thresholds when an initial force-constant model is available. Such randomized perturbation strategies have recently proven highly effective in identifying hidden, lower-symmetry ground states in other complex framework materials~\cite{zhu+24npjcmat,elen+25npjcmat}. Since stochastic initializations sample the PES non-deterministically, the resulting relaxed minima depend on the chosen displacement amplitude, random seed, cell-volume constraints, and minimization algorithm. Consequently, any low-energy configuration identified through stochastic rattling requires specific validation: the relaxation trajectory must be reproducible across independent random seeds, and the recomputed phonon dispersion of the final symmetry-lowered phase must be confirmed free of residual imaginary branches. For MOF-5, explicit one-dimensional mode mapping (Sec.~\ref{sec:modemapping}) successfully resolved the soft-mode instability, making stochastic rattling unnecessary. However, for frameworks exhibiting dense networks of coupled soft modes, stochastic rattling represents the logical next step in the diagnostic hierarchy.

Beyond static $T=0$ PES mapping, a complete picture of framework stability often requires evaluating thermal fluctuations. The complementary roles of static and dynamic approaches can be classified into three distinct regimes. The harmonic mode tracking ($T=0~\mathrm{K}$) explicitly identifies local saddle points, unstable normal-mode eigenvectors, and exact energy gains in the immediate vicinity of the reference structure. The stochastic structure searching, again at $T=0~\mathrm{K}$, explores distant, highly coupled, lower-symmetry minima across complex energy landscapes when single-mode tracks prove inconclusive. Finally, finite-temperature sampling captures dynamic disorder and thermal populations via finite-temperature MC or molecular dynamics simulations. Finite-temperature methods are particularly relevant for frameworks containing molecular rotors, where collective dipole ordering or rotational phase transitions emerge from coupled local degrees of freedom~\cite{su+21natchem, pere+23angew, berg+26jcp}. In such systems, dynamic disorder dominates, and a single static ground-state structure no longer provides a physically meaningful description. Matching the diagnostic strategy to the underlying physics of the framework ensures a computationally efficient path to accurate lattice dynamics.

\section{Summary and conclusions}
In this work, we established a diagnostic protocol to distinguish numerical artifacts from intrinsic structural instabilities in MOFs, using MOF-5 as a representative example. Our analysis shows that the low-frequency phonon spectrum of this framework is highly sensitive not only to the choice of the xc functional and dispersion correction scheme, but also, and often more critically, to numerical force thresholds, grid resolutions, symmetry-standardization tolerances, and the simulation cell representation. Crucially, standard settings that are considered robust for electronic-structure calculations prove insufficient for lattice dynamics, frequently introducing severe numerical noise that manifests as spurious imaginary modes. 

Quantitatively, we found that numerical artifacts span minimum frequencies from $\nu_{\min}\approx-0.3$ to $-0.8$~THz for grid- or force-induced noise, and reach $-4.7$~THz for under-converged geometries for which symmetry standardization yields a fragmented primitive-cell representation. Across all cases, residual acoustic-sum-rule violation remains negligible ($\sim10^{-6}$~THz), justifying the use of the minimum frequency $\nu_{\min}$ as a quantitative descriptor for dynamical stability: a soft mode is classified as a numerical artifact when $\nu_{\min}$ is restored to zero by tightening a single convergence parameter, and as intrinsic only when it survives the complete multi-parameter screening, as for the residual $-0.44$~THz $A_{2g}$ mode of the high-symmetry MOF-5 phase. 

To mitigate these issues, we recommend adopting the conventional cubic cell as the standard reference configuration for production-level phonon calculations in MOF-5 and analogous architectures. This representation avoids the ambiguity in coordinate wrapping that can obscure under-converged relaxations in primitive cells, and reduces the workflow sensitivity to symmetry-standardization tolerances, establishing a stable basis for interatomic force constant evaluation. Remarkably, this conventional-cell approach offers a 60\% reduction in total computational wall time compared to standard $2\times2\times2$ primitive supercell expansions without sacrificing spectral accuracy. Once this technical screening is completed, any remaining imaginary frequencies can be confidently assigned a physical interpretation. For MOF-5, mode mapping along the unstable $A_{2g}$ normal coordinate reveals a shallow displacive pathway connecting the high-symmetry $Fm\bar{3}m$ phase to a stable, symmetry-lowered $Fm\bar{3}$ local minimum. For more complex landscapes where multiple unstable branches coexist, stochastic Monte Carlo and phonon-rattle schemes represent the logical next step for breaking symmetry and mapping highly coupled PESs.

More broadly, this diagnostic protocol provides a reliable roadmap for predicting the structural stability of other flexible porous materials from first principles. Our findings suggest that high-throughput phonon workflows optimized for rigid, closed-packed inorganic crystals are not straightforwardly transferable to MOFs. Instead, the intrinsic softness of these porous materials demands a careful, system-specific validation of numerical convergence. Ultimately, this work suggests that the standard harmonic approximation and the computational protocols for phonon calculations originally designed for rigid, covalent semiconductors are reaching their limits when applied to the complex, flatter potential-energy landscapes of framework materials. As the field moves toward increasingly flexible and multi-functional building blocks, the development of advanced methods that explicitly capture higher-order anharmonicity and large-amplitude collective motions will be essential for the predictive modeling of framework stability.

\ack{J.S.A. thanks Jos\'e J. Plata for helpful discussions and feedback, and acknowledges funding from the Evonik Stiftung. This work was partly funded by the German Federal Ministry of Education and Research (Professorinnenprogramm III) and from the State of Lower Saxony (Professorinnen f\"ur Niedersachsen). The computational resources were provided by the North-German Supercomputing Alliance (NHR), project nic00084.}
\data{The data supporting the findings of this study are available in Zenodo at \url{https://doi.org/10.5281/zenodo.20340048}.}
\bibliographystyle{iopart-num}

\begin{thebibliography}{10}
\expandafter\ifx\csname url\endcsname\relax
  \def\url#1{{\tt #1}}\fi
\expandafter\ifx\csname urlprefix\endcsname\relax\def\urlprefix{URL }\fi
\providecommand{\eprint}[2][]{\url{#2}}
% Bibliography created with iopart-num v2.1
% /biblio/bibtex/contrib/iopart-num

\bibitem{yagh+03nature}
Yaghi O~M, O'Keeffe M, Ockwig N~W, Chae H~K, Eddaoudi M and Kim J 2003 {\em Nature\/} {\bf 423} 705--714 \urlprefix\url{https://doi.org/10.1038/nature01650}

\bibitem{kalm+18sciadv}
Kalmutzki M~J, Hanikel N and Yaghi O~M 2018 {\em Sci.~Adv.\/} {\bf 4} eaat9180 \urlprefix\url{https://doi.org/10.1126/sciadv.aat9180}

\bibitem{chen+22acr}
Chen Z, Kirlikovali K~O, Li P and Farha O~K 2022 {\em Acc.~Chem.~Res.\/} {\bf 55} 579--591 \urlprefix\url{https://doi.org/10.1021/acs.accounts.1c00707}

\bibitem{sumi+12cr}
Sumida K, Rogow D~L, Mason J~A, McDonald T~M, Bloch E~D, Herm Z~R, Bae T~H and Long J~R 2012 {\em Chem.~Rev.\/} {\bf 112} 724--781 \urlprefix\url{https://doi.org/10.1021/cr2003272}

\bibitem{kneb+22nnano}
Knebel A and Caro J 2022 {\em Nat.~Nanotechnol.\/} {\bf 17} 911--923 \urlprefix\url{https://doi.org/10.1038/s41565-022-01168-3}

\bibitem{li+99nature}
Li H, Eddaoudi M, O'Keeffe M and Yaghi O~M 1999 {\em Nature\/} {\bf 402} 276--279 ISSN 1476-4687 \urlprefix\url{https://doi.org/10.1038/46248}

\bibitem{edda+00jacs}
Eddaoudi M, Li H and Yaghi O~M 2000 {\em J.~Am.~Chem.~Soc.\/} {\bf 122} 1391--1397 \urlprefix\url{https://doi.org/10.1021/ja9933386}

\bibitem{lock+13dt}
Lock N, Christensen M, Wu Y, Peterson V~K, Thomsen M~K, Piltz R~O, Ramirez-Cuesta A~J, McIntyre G~J, Nor{\'e}n K, Kutteh R, Kepert C~J, Kearley G~J and Iversen B~B 2013 {\em Dalton~Trans.\/} {\bf 42} 1996--2007 \urlprefix\url{https://doi.org/10.1039/C2DT31491F}

\bibitem{rimm+14pccp}
Rimmer L~H~N, Dove M~T, Goodwin A~L and Palmer D~C 2014 {\em Phys.~Chem.~Chem.~Phys.\/} {\bf 16} 21144--21152 \urlprefix\url{https://doi.org/10.1039/C4CP01701C}

\bibitem{ryde+19adts}
Ryder M~R, Maul J, Civalleri B and Erba A 2019 {\em Adv.~Theory~Simul.\/} {\bf 2} 1900093 \urlprefix\url{https://doi.org/10.1002/adts.201900093}

\bibitem{edza+25jctc}
Edzards J, Santana-Andreo J, Saßnick H~D and Cocchi C 2025 {\em J.~Chem.~Theory.~Comput.\/} {\bf 21} 7062--7074 pMID: 40653644 \urlprefix\url{https://doi.org/10.1021/acs.jctc.5c00399}

\bibitem{andr+20jacs}
Andreeva A~B, Le K~N, Chen L, Kellman M~E, Hendon C~H and Brozek C~K 2020 {\em J.~Am.~Chem.~Soc.\/} {\bf 142} 19291--19299 \urlprefix\url{https://doi.org/10.1021/jacs.0c09499}

\bibitem{dubb+07jpcc}
Dubbeldam D, Walton K~S, Ellis D~E and Snurr R~Q 2007 {\em Angew.~Chem.~Int.~Ed.\/} {\bf 46} 4496--4499 \urlprefix\url{https://doi.org/10.1002/anie.200700218}

\bibitem{baro+01rmp}
Baroni S, de~Gironcoli S, Dal~Corso A and Giannozzi P 2001 {\em Rev.~Mod.~Phys.\/} {\bf 73} 515--562 \urlprefix\url{https://doi.org/10.1103/RevModPhys.73.515}

\bibitem{pall+22es}
Pallikara I, Kayastha P, Skelton J~M and Whalley L~D 2022 {\em Electron.~Struct.\/} {\bf 4} 033002 \urlprefix\url{https://doi.org/10.1088/2516-1075/ac78b3}

\bibitem{kama+26arxiv}
Kamath P~D and Persson K~A 2026 {\em arXiv preprint arXiv:2602.07295\/} \urlprefix\url{https://arxiv.org/abs/2602.07295}

\bibitem{togo+23jpsj}
Togo A 2023 {\em J. Phys. Soc. Jpn.\/} {\bf 92} 012001 \urlprefix\url{https://doi.org/10.7566/JPSJ.92.012001}

\bibitem{zhu+24npjcmat}
Zhu Z, Park J, Sahasrabuddhe H, Ganose A~M, Chang R, Lawson J~W and Jain A 2024 {\em npj~Comp.~Mater.\/} {\bf 10} 258 \urlprefix\url{https://doi.org/10.1038/s41524-024-01437-w}

\bibitem{hohe+64pr}
Hohenberg P and Kohn W 1964 {\em Phys.~Rev.\/} {\bf 136}(3B) B864--B871 \urlprefix\url{https://link.aps.org/doi/10.1103/PhysRev.136.B864}

\bibitem{kohn+65pr}
Kohn W and Sham L~J 1965 {\em Phys.~Rev.\/} {\bf 140}(4A) A1133--A1138 \urlprefix\url{https://link.aps.org/doi/10.1103/PhysRev.140.A1133}

\bibitem{perd+96prl}
Perdew J~P, Burke K and Ernzerhof M 1996 {\em Phys.~Rev.~Lett.\/} {\bf 77}(18) 3865--3868 \urlprefix\url{https://doi.org/10.1103/PhysRevLett.77.3865}

\bibitem{furn+20jpcl}
Furness J~W, Kaplan A~D, Ning J, Perdew J~P and Sun J 2020 {\em J.~Phys.~Chem.~Lett.\/} {\bf 11} 8208--8215 \urlprefix\url{https://doi.org/10.1021/acs.jpclett.0c02405}

\bibitem{grim+10jcp}
Grimme S, Antony J, Ehrlich S and Krieg H 2010 {\em J.~Chem.~Phys.\/} {\bf 132} 154104 ISSN 0021-9606 \urlprefix\url{https://doi.org/10.1063/1.3382344}

\bibitem{grim+11jcc}
Grimme S, Ehrlich S and Goerigk L 2011 {\em J. Comput. Chem.\/} {\bf 32} 1456--1465 \urlprefix\url{https://doi.org/10.1002/jcc.21759}

\bibitem{saba+13prb}
Sabatini R, Gorni T and de~Gironcoli S 2013 {\em Phys.~Rev.~B\/} {\bf 87}(4) 041108 \urlprefix\url{https://link.aps.org/doi/10.1103/PhysRevB.87.041108}

\bibitem{born+54dtcl}
Born M and Huang K 1954 {\em Dynamical Theory of Crystal Lattices\/} (Oxford: Clarendon Press) \urlprefix\url{https://doi.org/10.1093/oso/9780192670083.001.0001}

\bibitem{togo+15scripta}
Togo A and Tanaka I 2015 {\em Scripta Mater.\/} {\bf 108} 1--5 \urlprefix\url{https://doi.org/10.1016/j.scriptamat.2015.07.021}

\bibitem{parl+97prl}
Parlinski K, Li Z~Q and Kawazoe Y 1997 {\em Phys. Rev. Lett.\/} {\bf 78} 4063--4066 \urlprefix\url{https://doi.org/10.1103/PhysRevLett.78.4063}

\bibitem{hoff+22jmca}
Hoffman A~E~J, Senkovska I, Wieme J, Krylov A, Kaskel S and Van~Speybroeck V 2022 {\em J.~Mater.~Chem.~A\/} {\bf 10} 17254--17266 \urlprefix\url{https://doi.org/10.1039/D2TA01678H}

\bibitem{kuch+19zkri}
Kuchta B, Formalik F, Rogacka J, Neimark A~V and Firlej L 2019 {\em Z.~Phys.\/} {\bf 234} 513--527 \urlprefix\url{https://doi.org/10.1515/zkri-2018-2152}

\bibitem{kuhn+20jcp}
K{\"u}hne T~D, Iannuzzi M, Del~Ben M, Rybkin V~V, Hutter J {\em et~al.\/} 2020 {\em J. Chem. Phys.\/} {\bf 152} 194103 \urlprefix\url{https://doi.org/10.1063/5.0007045}

\bibitem{goed+96prb}
Goedecker S, Teter M and Hutter J 1996 {\em Phys.~Rev.~B\/} {\bf 54}(3) 1703--1710 \urlprefix\url{https://link.aps.org/doi/10.1103/PhysRevB.54.1703}

\bibitem{vand-hutt07jcp}
VandeVondele J and Hutter J 2007 {\em J.~Chem.~Phys.\/} {\bf 127} 114105 ISSN 0021-9606 (\textit{Preprint} \eprint{https://pubs.aip.org/aip/jcp/article-pdf/doi/10.1063/1.2770708/14787836/114105_1_online.pdf}) \urlprefix\url{https://doi.org/10.1063/1.2770708}

\bibitem{lipp+97mp}
Lippert G, Hutter J and Parrinello M 1997 {\em Mol.~Phys.\/} {\bf 92} 477--488 \urlprefix\url{https://doi.org/10.1080/002689797170220}

\bibitem{vand+05cpc}
VandeVondele J, Krack M, Mohamed F, Parrinello M, Chassaing T and Hutter J 2005 {\em Comput. Phys. Commun.\/} {\bf 167} 103--128 \urlprefix\url{https://doi.org/10.1016/j.cpc.2004.12.014}

\bibitem{skel+17git}
Skelton J~M 2017 Modemap \url{https://github.com/JMSkelton/ModeMap} accessed: 2026-04-26 \urlprefix\url{https://github.com/JMSkelton/ModeMap}

\bibitem{sety+10cms}
Setyawan W and Curtarolo S 2010 {\em Comp.~Mater.~Sci.\/} {\bf 49} 299--312 \urlprefix\url{https://doi.org/10.1016/j.commatsci.2010.05.010}

\bibitem{sass+24aim2dat}
Saßnick H~D, Edzards J, Reents T and Cocchi C 2026 {\em Electron.~Struct.\/} {\bf 8} 037001 \urlprefix\url{https://doi.org/10.1088/2516-1075/ae8964}

\bibitem{adam+99jcp}
Adamo C and Barone V 1999 {\em J.~Chem.~Phys.\/} {\bf 110} 6158--6170 ISSN 0021-9606 (\textit{Preprint} \eprint{https://pubs.aip.org/aip/jcp/article-pdf/110/13/6158/19068890/6158\_1\_online.pdf}) \urlprefix\url{https://doi.org/10.1063/1.478522}

\bibitem{sunj+15prl}
Sun J, Ruzsinszky A and Perdew J~P 2015 {\em Phys.~Rev.~Lett.\/} {\bf 115} 036402 \urlprefix\url{https://doi.org/10.1103/PhysRevLett.115.036402}

\bibitem{ning+22prb}
Ning J, Kothakonda M, Furness J~W, Kaplan A~D, Ehlert S, Brandenburg J~G, Perdew J~P and Sun J 2022 {\em Phys.~Rev.~B\/} {\bf 106}(7) 075422 \urlprefix\url{https://link.aps.org/doi/10.1103/PhysRevB.106.075422}

\bibitem{tan+11csrv}
Tan J~C and Cheetham A~K 2011 {\em Chem.~Soc.~Rev.\/} {\bf 40} 1059--1080 \urlprefix\url{https://doi.org/10.1039/C0CS00163E}

\bibitem{redf+19csci}
Redfern L~R and Farha O~K 2019 {\em Chem.~Sci.\/} {\bf 10} 10666--10679 \urlprefix\url{https://doi.org/10.1039/C9SC04249K}

\bibitem{krau+20anie}
Krause S, Hosono N and Kitagawa S 2020 {\em Angew.~Chem.~Int.~Ed.\/} {\bf 59} 15325--15341 \urlprefix\url{https://doi.org/10.1002/anie.202004535}

\bibitem{gan+21cpc}
Gan C~K, Liu Y, Sum T~C and Hippalgaonkar K 2021 {\em Comput. Phys. Commun.\/} {\bf 259} 107635

\bibitem{togo+24stamm}
Togo A, Shinohara K and Tanaka I 2024 {\em Sci.~Technol.~Adv.~Mater.~Meth.\/} {\bf 4} 2384822--2384836 \urlprefix\url{https://doi.org/10.1080/27660400.2024.2384822}

\bibitem{skel+16prl}
Skelton J~M, Burton L~A, Parker S~C, Walsh A, Kim C~E, Soon A, Buckeridge J, Sokol A~A, Catlow C~R~A, Togo A and Tanaka I 2016 {\em Phys.~Rev.~Lett.\/} {\bf 117} 075502 \urlprefix\url{https://doi.org/10.1103/PhysRevLett.117.075502}

\bibitem{whal+16prb}
Whalley L~D, Skelton J~M, Frost J~M and Walsh A 2016 {\em Phys.~Rev.~B\/} {\bf 94} 220301 \urlprefix\url{https://doi.org/10.1103/PhysRevB.94.220301}

\bibitem{bhog-wagh21jpcc}
Bhogra M and Waghmare U~V 2021 {\em J.~Phys.~Chem.~C\/} {\bf 125} 14924--14931 \urlprefix\url{https://doi.org/10.1021/acs.jpcc.1c02598}

\bibitem{erik+19adts}
Eriksson F, Fransson E and Erhart P 2019 {\em Adv.~Theory Simul.\/} {\bf 2} 1800184 \urlprefix\url{https://doi.org/10.1002/adts.201800184}

\bibitem{west+06prl}
West D and Estreicher S~K 2006 {\em Phys.~Rev.~Lett.\/} {\bf 96} 115504 \urlprefix\url{https://doi.org/10.1103/PhysRevLett.96.115504}

\bibitem{elen+25npjcmat}
Elena A~M, Kamath P~D, Jaffrelot~Inizan T, Rosen A~S, Zanca F {\em et~al.\/} 2025 {\em npj~Comp.~Mater.\/} {\bf 11} 125 \urlprefix\url{https://doi.org/10.1038/s41524-025-01611-8}

\bibitem{su+21natchem}
Su Y~S {\em et~al.\/} 2021 {\em Nat. Chem.\/} {\bf 13} 278--283

\bibitem{pere+23angew}
Perego J {\em et~al.\/} 2023 {\em Angew. Chem. Int. Ed.\/} {\bf 62} e202215893

\bibitem{berg+26jcp}
Bergler T, Badalov S, Wixforth A, Volkmer D and Oberhofer H 2026 {\em J. Chem. Phys.\/} {\bf 164} 044125

\end{thebibliography}
\providecommand{\newblock}{}

\end{document}